\documentclass{webofc}
\usepackage[varg]{txfonts}   
\newcommand{\pt}{$p_{\rm{T}}$}

\newcommand{\detadphi}{$\Delta\eta\Delta\varphi$}
\newcommand{\ee}{$e^+e^-$}
\begin{document}
\title{Insight into particle production mechanisms via angular correlations of identified particles measured with ALICE in pp collisions at  $\mathbf{\sqrt{{\textit s}}}=7$~TeV}
%
%

\author{\firstname{Ma{\l}gorzata Anna} \lastname{Janik}\inst{1}\fnsep\thanks{\email{majanik@if.pw.edu.pl}} 
}

\institute{Faculty of Physics, Warsaw University of Technology, Poland}

\abstract{%
Two-particle correlations as a function of $\Delta\eta$ and $\Delta\varphi$ are used in all collision systems to study a wide range of physical phenomena. Examples include the collective behavior of the quark-gluon plasma medium, jets, quantum statistics or Coulomb effects, conservation laws, and resonance decays.
In this work, measurements of the correlations of identified particles and their antiparticles (for $\pi$, K, p, $\Lambda$) are reported in pp collisions at $\sqrt{s}=7$~TeV at low transverse momenta. The analysis reveals differences in particle production between baryons and mesons. The correlation functions for mesons exhibit the expected peak dominated by effects of mini-jet fragmentation and are reproduced well by general purpose Monte Carlo generators. For baryon pairs where both particles have the same baryon number, a near-side anti-correlation structure is observed instead of a peak; our experimental observation present a challenge to the contemporary models (PYTHIA, PHOJET). This effect is further interpreted in the context of baryon production mechanisms in the fragmentation processes.
}
\maketitle
\section{Introduction}
\label{intro}

Studies of particle production mechanisms in elementary collisions date back to the times of R. Feynman and R. Field, who proposed in 1977 a simple mechanism describing the principles of creation of the so-called ``jets'', collimated streams of particles \cite{Field:1977fa}. They proposed rules on how the particles are created, how the energy is distributed and considered limitations connected to the conservation laws.
Elements of the proposed scheme are used even today in the most popular fragmentation models (such as ``Lund model'' employed in the PYTHIA generator). However, the implementation details have to be compatible with the experimental data. It is then the task for the experiment to provide basic information: How strong should be the correlations between created hadrons? How does this correlation change, when two or more baryons or strange particles are created? Answers to these and other questions have been searched so far only in \ee\ collisions, at much lower energies than those achieved by contemporary high energy physics experiments and on substantially smaller data samples \cite{Aihara:1986fy,Acton:1993ux}. 

We address the above questions using two-particle angular correlations. This tool allows for a broad exploration of the underlying physics phenomena of particle production in	collisions of both protons and heavy ions by measuring angular distributions in \detadphi~space (where $\Delta\eta$ is the pseudorapidity difference and $\Delta\varphi$ is the azimuthal angle difference between two particles).
The analysis is performed for identified particles,
	that is pions, kaons, protons, and lambda particles, produced in pp collisions at $\sqrt{s}=7$~TeV recorded by the ALICE detector in 2010~\cite{Adam:2016iwf}. 
The measured correlations should be sensitive to conservation laws as well as details of
	particle production mechanisms, including the parton fragmentation. In order to interpret the data in this context, dedicated Monte Carlo simulations were performed. 

\section{Analysis}
\label{sec:data}
The studies were done separately for particle--particle and anti-particle--anti-particle pairs, and for four particle species ($\pi$, K, p, $\Lambda$) \cite{Adam:2016iwf}. All particles used for the analysis were measured at low transverse momenta, up to 2.5 GeV/$c$, within the pseudorapidity range $|\eta|<0.8$. 
The particle identification of pions, kaons, and protons was performed on a track-by-track basis using information from the Time Projection Chamber and Time-Of-Flight detectors\cite{Aamodt:2008zz}. The applied procedure resulted in a purity above 99\% for pions and protons and above 96\% for kaons. The lambda baryons were reconstructed using
their distinctive decay topology in the channel $\Lambda(\overline{\Lambda})\rightarrow\mathrm{p}\pi^-(\overline{\mathrm{p}}\pi^+)$. The $\Lambda$ purity (defined as $S$/$(S+B)$) was found to be above 95\%~\cite{Adam:2016iwf}.

The reported experimental correlation function is constructed as 
\begin{equation}
\label{eq:CorrelationFuntion}
C(\Delta\eta,\Delta\varphi)=\frac{S(\Delta\eta,\Delta\varphi)}{B(\Delta\eta,\Delta\varphi)},
\end{equation}
where $\Delta\eta=\eta_1 - \eta_2$ is the difference in
pseudorapidity, $\Delta\varphi=\varphi_1 - \varphi_2$ is the
difference in azimuthal angle. $S(\Delta\eta,\Delta\varphi)$ is the distribution of correlated pairs and $B(\Delta\eta,\Delta\varphi)$ is the reference distribution, calculated using the mixed-events technique, reflecting the single-particle
acceptance. 
Both the $S$ and $B$ distributions are normalized by the respective number of pairs, therefore, the reported distribution is a ratio of probabilities.\footnote{A conditional probability to observe a particle with azimuthal angle $\varphi_1$ and pseudorapidity $\eta_1$ if a particle with azimuthal angle $\varphi_2$ and psedurapidity $\eta_2 $ is observed as well. In the absence of correlations, the ratio should equal unity.} All details of the analysis can be found in ~\cite{Adam:2016iwf}.

\section{Results}
\label{sec:results}

\begin{figure}[h]
	\centering
	\includegraphics[width=0.76\textwidth]{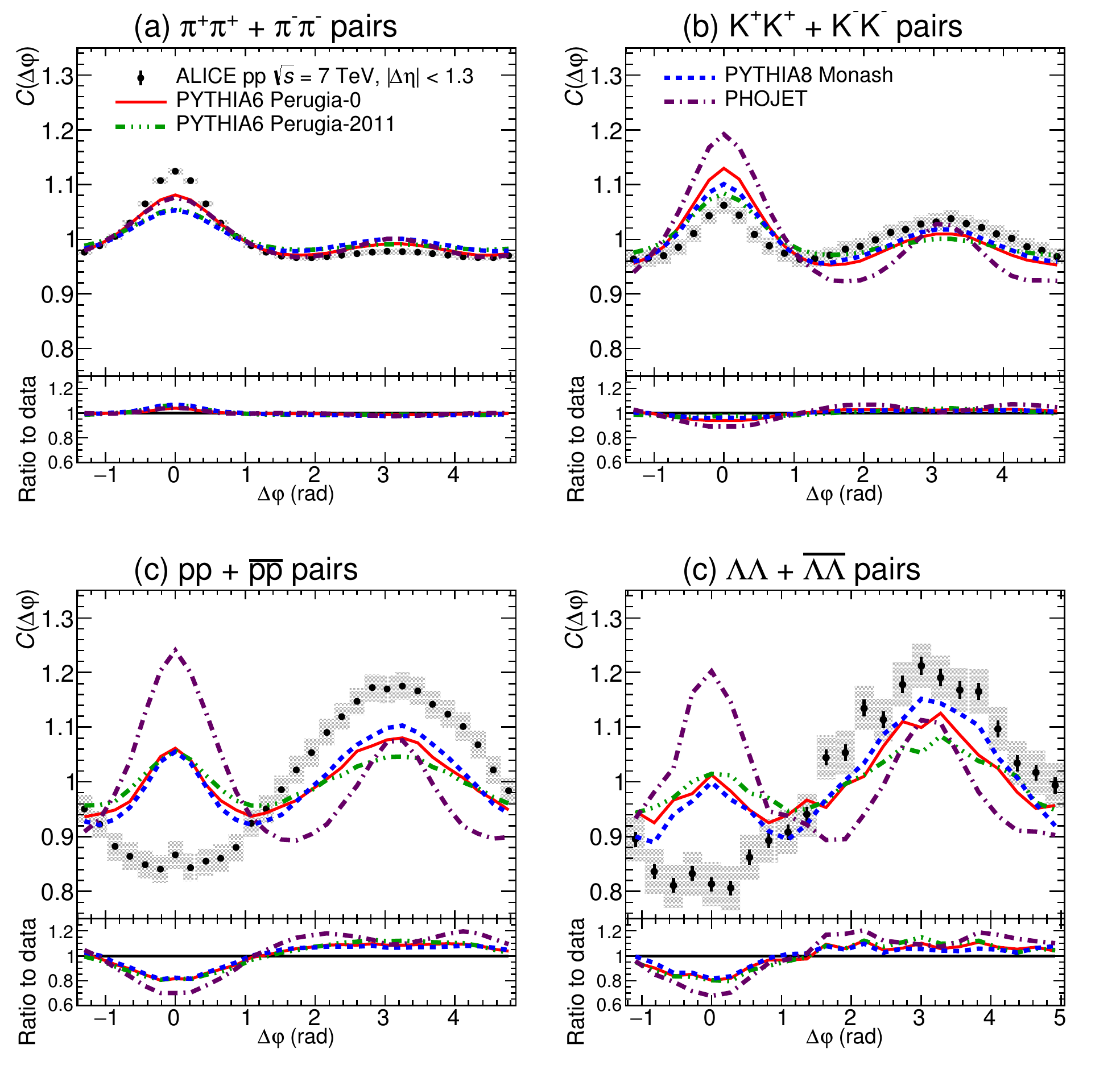}
	\caption{Projections of correlation functions integrated over $\Delta\eta$ for combined pairs of (a) $\pi^+\pi^+ + \pi^-\pi^-$, (b) $\rm K^+K^+ + K^-K^-$, (c) $\mathrm{pp} + \rm \overline{p}\overline{p}$ and (d) $\Lambda\Lambda+\overline{\Lambda}\overline{\Lambda}$, obtained from ALICE data and four Monte Carlo models (PYTHIA6 Perugia-0, PYTHIA6 Perugia-2011, PYTHIA8 Monash, PHOJET). Bottom panels show ratios of MC models to ALICE data. Statistical (bars) and systematic (boxes) uncertainties are plotted.  Plot from \cite{Adam:2016iwf}.}
	\label{Fig:Proj_CompMC_PIpPIp_KpKp_PP}
\end{figure}

In Fig.~\ref{Fig:Proj_CompMC_PIpPIp_KpKp_PP} correlation functions are presented for like-sign pairs for (a) pions, (b) kaons, (c) protons, and (d) lambdas.\footnote{For the full collection of results including two-dimensional \detadphi\ correlations see \cite{Adam:2016iwf}.} 
Results show significantly different behavior of identical meson and baryon pairs: mesons (pions and kaons) exhibit a near-side peak  for $\Delta\varphi = 0$, while for baryons (p and $\Lambda$) a depression is observed in this region.

The correlation functions are compared to predictions of Monte Carlo (MC) models. The following MC event generators were used: PYTHIA6.4 tunes Perugia-0 and Perugia-2011~\cite{Sjostrand:2006za,Skands:2010ak}, PYTHIA8 Monash
tune~\cite{Sjostrand:2007gs,Skands:2014pea}, and PHOJET version 1.12~\cite{Engel:1994vs}.
The correlations between mesons are qualitatively reflected by the models (the difference visible for pions comes from Bose--Einstein correlations which are absent in the studied MC samples).
However, for baryon pairs significant differences can be seen, that are not only quantitative, but qualitative. Models fail to reproduce the depression visible for like-sign baryons. All studied generators frequently produce two baryons close in
phase-space (e.g. within the mini-jet peak), which is not reflected in the experimental data.  Further studies performed with EPOS-LHC \cite{Pierog:2013ria} and HERWIG \cite{Corcella:2000bw} models  (not shown on the plot) do not differ qualitatively from PYTHIA and PHOJET.

\begin{figure}[ht]
	\centering
	\includegraphics[width=0.45\textwidth]{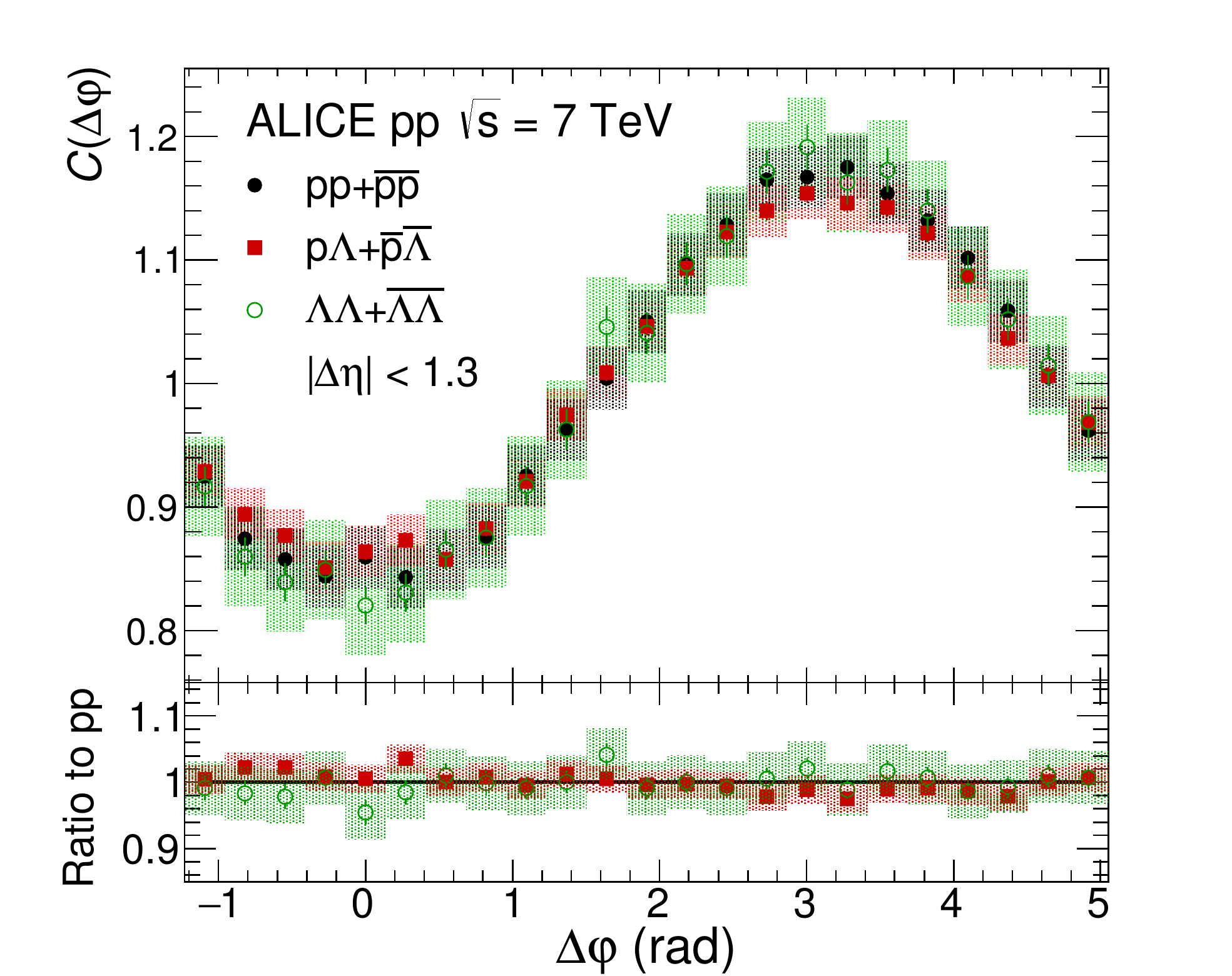}
	\includegraphics[width=0.45\textwidth]{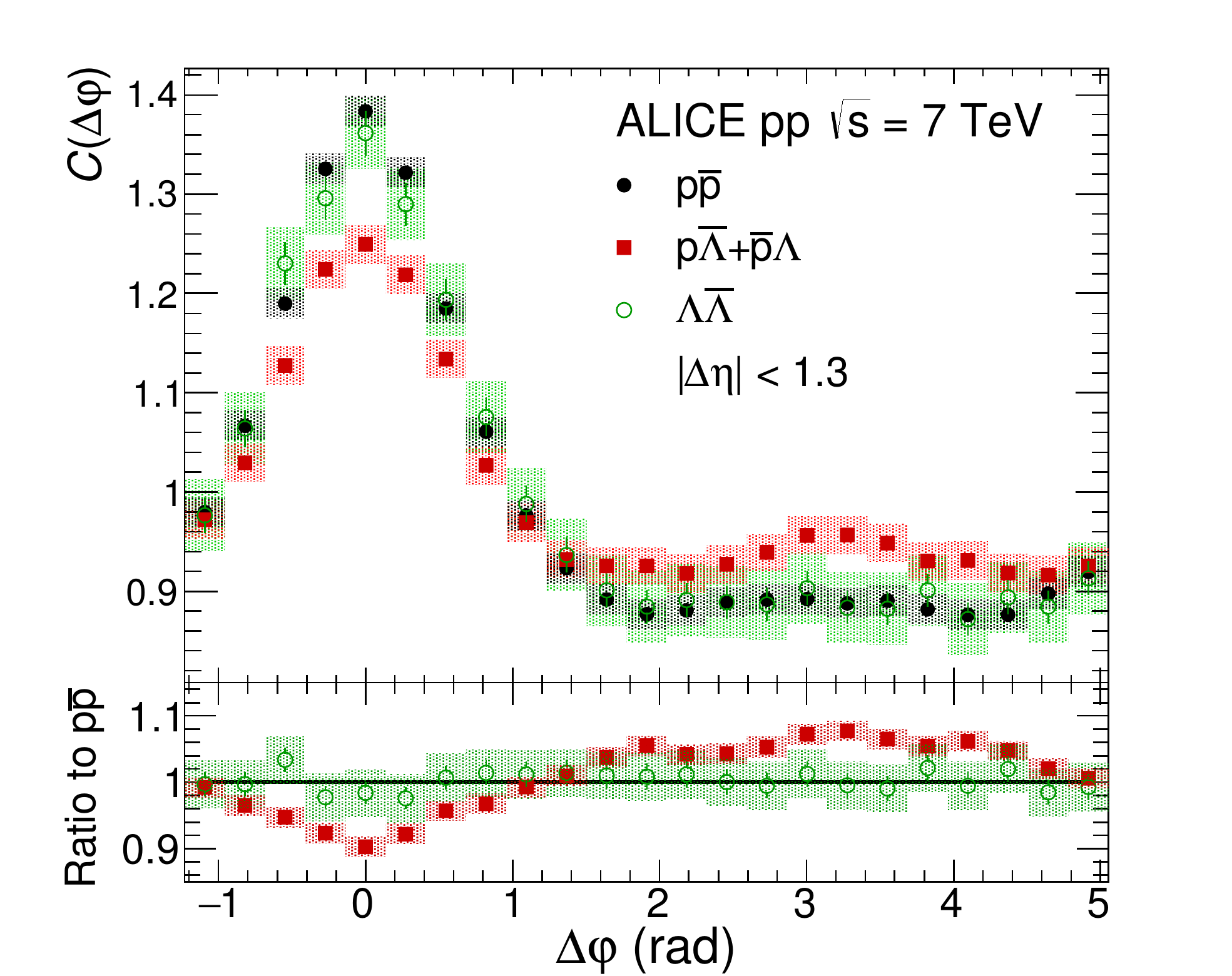}
	\caption{Projections of correlation functions integrated over $\Delta\eta$ for combined pairs of (left)  $\rm pp+\overline{p}\overline{p}$, $\rm p\Lambda+\overline{p}\overline{\Lambda}$, and $\Lambda\Lambda+\overline{\Lambda}\overline{\Lambda}$  and (right) $\rm p\overline{p}$, $\rm p\overline{\Lambda}+\overline{p}\Lambda$, and $\Lambda\overline{\Lambda}$. Statistical (bars) and systematic (boxes) uncertainties are plotted. Plot from \cite{Adam:2016iwf}.
	}
	\label{Fig:Proj_PAP_PAL_LAL_LHC10bcde}
\end{figure}

We studied if some well-known physics effects, not included in the models, can influence the shape of baryon-baryon correlation functions. We performed the comparison of all baryon pairs, which can be seen in Fig.~\ref{Fig:Proj_PAP_PAL_LAL_LHC10bcde}. We observe that:
\begin{itemize}
	\item The Coulomb effect plays a marginal role: the shape of the correlation function for all studied baryon--baryon (and baryon--anti-baryon) pairs is  similar, regardless of the electric charge of the particles. 
	\item Fermi-Dirac quantum statistics is not the cause of the observed depression: the same magnitude of depression is observed for pp, $\Lambda \Lambda$ (identical particles), and p$\Lambda$ (non-identical particles).
	\item Local baryon number conservation is not the only source of the depression. All studied models include this mechanism, but still are not able to reproduce the experimental data.
\end{itemize}

\clearpage

\vspace{-0.9cm}
\section{Summary}
\label{sec:summary}

Angular correlations of identified particles ($\pi$, K, p, $\Lambda$) were analyzed in pp collisions at $\sqrt{s}=7$~TeV recorded with the ALICE experiment. A significant depression at $(\Delta\eta,\Delta\varphi)\approx(0,0)$ is observed for the baryon--baryon and anti-baryon--anti-baryon pairs, which is not seen for mesons nor for baryon--anti-baryon pairs. This depression is not reproduced by Monte Carlo models. This suggests that (a) the jet fragmentation is not the dominant mechanism involved in the production of baryons in the studied \pt\ range as models indicate, or (b) the fragmentation mechanisms employed in PYTHIA and PHOJET are incomplete. The latter scenario would further suggest that some additional, not identified mechanism must exist. Such mechanism would suppress the production of more than one baryon--anti-baryon pair during a single fragmentation. Therefore, the presented results may suggest the need to modify particle production mechanisms as well as the modification of fragmentation functions in models.

\vspace{-0.2cm}
\section*{Acknowledgements}
\vspace{-0.1cm}
This work was supported by the Polish National Science Centre under decisions no. 2014/13/B/ST2/04054, no. 2015/19/D/ST2/01600 and no. 2016/22/M/ST2/00176.
\vspace{-0.3cm}
\bibliography{biblio}

\end{document}